\documentclass[twocolumn,showpacs,preprintnumbers,nofootinbib,amsmath,amssymb,floatfix,aps,prc]{revtex4-1}
\usepackage{graphicx}
\usepackage{isotope}
\usepackage{bm}

\newcommand{\qrec}{\ensuremath{Q_{\textrm{rec}}^2}}

\begin{document}

\title{Charged-current quasielastic scattering of muon antineutrino and neutrino\\ in the MINERvA experiment}
\author{Artur M. Ankowski}
\email{ankowski@vt.edu}
\affiliation{Center for Neutrino Physics, Virginia Tech, Blacksburg, Virginia 24061, USA}

\date{\today}%

\begin{abstract}
One of the largest sources of systematic uncertainties in ongoing neutrino-oscillation measurements is the description of nuclear effects. Its considerable reduction is expected thanks to the dedicated studies of (anti)neutrino-nucleus interactions in the MINERvA experiment. In this article, the calculations within the spectral function approach are compared to the charged-current quasielastic cross sections reported from MINERvA. The obtained results show that the effect of final-state interactions on the (anti)muon kinematics plays a pivotal role in reproducing the experimental data.
\end{abstract}


\maketitle

The MINERvA Collaboration has recently reported the flux-averaged differential cross section $d\sigma/d\qrec$, $\qrec$ being the reconstructed four-momentum transfer squared, for charged-current (CC) quasielastic (QE) scattering of muon antineutrinos~\cite{ref:MINERvA_anu} on nucleons in the hydrocarbon target, CH. In a subsequent paper~\cite{ref:MINERvA_nu}, the corresponding result for muon neutrinos has also been presented. While five theoretical calculations have been compared to the data in Refs.~\cite{ref:MINERvA_anu,ref:MINERvA_nu}, none of them turned out to be able to satisfactorily describe both the antineutrino and neutrino cross sections. Rather consistent results have been found using superscaling approaches~\cite{ref:SuSA_MINERvA,ref:SuSAv2_MINERvA,ref:SuSAv2&MEC_MINERvA} and the local Fermi gas model with effective interactions and the random-phase approximation effects~\cite{ref:Nieves_MINERvA}.

In this article, I argue that the effects of final-state interactions (FSI) on the $\mu^\pm$ kinematics are essential to reproduce the MINERvA CC QE data, half of which correspond to $\qrec\leq0.2$ GeV$^2$, where the contribution of low momentum transfers is significant~\cite{ref:shape}.

Changing the energy balance in the primary vertex, interactions between the struck nucleon and the spectator
system affect the final energy of the charged lepton $E_\ell'$. Additional modifications to $E_\ell'$, different for $\ell^-$ and $\ell^+$, result from an influence of the Coulomb field of the nucleus on all charged particles in the final state~\cite{ref:EMA'}.

\begin{figure}[b]
\centering
    \includegraphics[width=0.8\columnwidth]{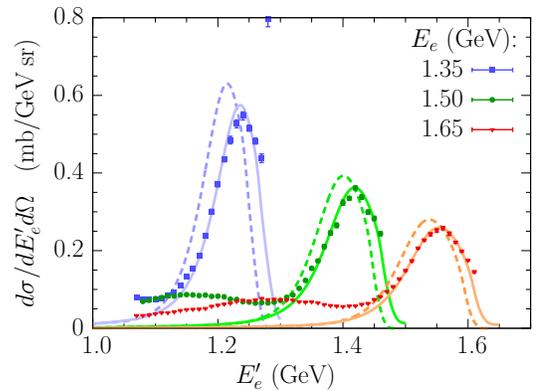}
\caption{\label{fig:electrons}(color online). Double differential $\isotope{C}(e,e')$ cross sections at scattering angle $13.5^\circ$ and beam energies 1.35, 1.50, and 1.65 GeV. The SF calculations without (dashed line) and with (solid lines) FSI effects~\cite{ref:FSI} are compared to the data~\cite{ref:Baran}.
}
\end{figure}

\begin{figure}
\centering
    \includegraphics[width=0.8\columnwidth]{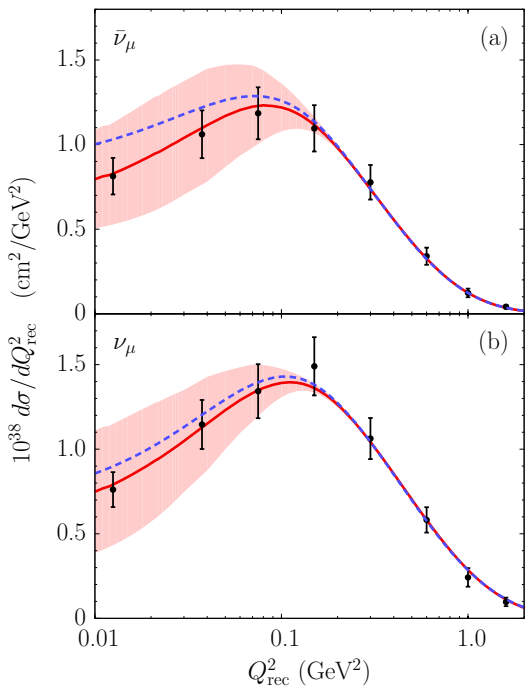}
\caption{\label{fig:neutrinos}(color online). Differential cross sections $d\sigma/d\qrec$
for CC QE (a) $\bar\nu_\mu$ and (b) $\nu_\mu$ scattering in MINERvA. The SF calculations without (dashed line) and with (solid lines) FSI effects~\cite{ref:FSI} are compared to the data~\cite{ref:MINERvA_anu,ref:MINERvA_nu}. The bands represent theoretical uncertainties.
}
\end{figure}

As known from electron scattering, FSI broaden the QE peak and shift it in a kinematics-dependent way, increasing typical $E_\ell'$ at low momentum transfers.  To illustrate those effects, in Fig.~\ref{fig:electrons} the predictions of the spectral function (SF) approach~\cite{ref:FSI}---involving no adjustable parameters---are compared to the experimental data of Ref.~\cite{ref:Baran}. The same mechanism redistributes the (anti)neutrino cross sections $d\sigma/d\qrec$ toward higher values of $\qrec$.

The role of FSI on electron and neutrino cross sections in the SF formalism has been studied in detail by the authors of Ref.~\cite{ref:FSI} for the carbon nucleus. In the approach of Ref.~\cite{ref:FSI}, the target's ground-state properties are described using the hole SF~\cite{ref:Omar_LDA}, consistently accounting for both the the shell structure~\cite{ref:Saclay_C, ref:Dutta} and nucleon-nucleon correlations~\cite{ref:Omar_NM}. To include the effect of Pauli blocking, the particle SF calculated following Ref.~\cite{ref:shape} is employed. Modification of the struck-particle's spectrum---due to its FSI with the spectator system---is accounted for in the energy conservation by making use of the Dirac phenomenological potential of Cooper {\it et al.}~\cite{ref:Cooper_EDAI}.
FSI-induced broadening of the QE peak is described within the correlated Glauber approximation~\cite{ref:Omar_FSI,ref:Omar_FSI_NM}, applying the nuclear transparency of carbon reported in Ref.~\cite{ref:Rohe}. It is noteworthy that a remarkable agreement with a large body of experimental $\isotope[12][6]{C}(e,e')$ data~\cite{ref:Barreau,ref:Whitney,ref:OConnell,ref:Baran,ref:Baghdasarian,ref:Sealock,ref:Day} has been observed in Ref.~\cite{ref:FSI}.


Within the approach of Ref.~\cite{ref:FSI}, I have calculated the CC QE $\bar\nu_\mu$ and $\nu_\mu$ cross sections averaged over the MINERvA fluxes, expressing both the carbon and free-proton (for $\bar\nu_\mu$ only) contributions as a function of $\qrec$ defined as in Refs.~\cite{ref:MINERvA_anu,ref:MINERvA_nu}. I have employed a conservative estimate of the nuclear model's uncertainties---represented by the bands in Fig.~\ref{fig:neutrinos}---coming from the optical potential parametrization and the contribution of the low-excitation energies $E_x\lesssim26$ MeV. Due to the stability of the carbon nucleus, I have assumed that the central values of the calculations correspond to $E_x\geq0$. To accommodate the contribution of two-body currents in an effective manner, the axial mass $M_A$ has been treated as a free parameter.

\begin{table}
\caption{\label{tab:fits} Fit results to the CC QE MINERvA data.}
\begin{ruledtabular}
\begin{tabular}{ c c c c }
 & antineutrino & neutrino & combined fit \\
\hline
 & \multicolumn{3}{c}{including theoretical uncertainties:}\\
$M_A$ (GeV) & $1.16\pm0.06$ & $1.17\pm0.06$ & $1.16\pm0.06$ \\
$\chi^2/$d.o.f. & 0.38  & 1.33  & 0.93 \\
\hline
 & \multicolumn{3}{c}{neglecting theoretical uncertainties:}\\
$M_A$ (GeV) & $1.15\pm0.10$ & $1.15\pm0.07$ & $1.13\pm0.06$ \\
$\chi^2/$d.o.f. & 0.44  & 1.38  & 1.00 \\
\hline
 & \multicolumn{3}{c}{neglecting FSI ($M_A=1.16$ GeV):}\\
$\chi^2/$d.o.f. & 2.49  & 2.45  & 2.42 \\
\end{tabular}
\end{ruledtabular}
\end{table}

In calculations of the $\chi^2$ distribution, I have accounted for correlations between the data, employing the standard expression
\[
\chi^2=\sum_{i,\,j}\frac{\Delta_i}{\sigma_i}V^{-1}_{ij}\frac{\Delta_j}{\sigma_j},
\]
where the sum runs over all data points, $\Delta_k$ is the difference between the theoretical and experimental value at point $k$, subject to the total uncertainty $\sigma_k$, and $V$ denotes the correlation matrix reported by the MINERvA Collaboration. Including
the theoretical uncertainty in $\sigma_k$, I have added it in quadrature with the experimental error.

The determined $M_A$ values and the corresponding $\chi^2$ per degree of freedom (d.o.f.) are given in Table~\ref{tab:fits}.
The difference with respect to $M_A=1.03$~GeV, extracted predominantly from the deuteron measurements~\cite{ref:Meissner}, indicates a nonvanishing contribution of reaction mechanisms other than those involving a one-body current.
Interestingly, the effective $M_A$ values turn out to be very consistent in the antineutrino and neutrino cases.

The analyses with and without theoretical uncertainties give rather consistent results for a twofold reason:
(i) theoretical uncertainties are only relevant at low $\qrec$, where the cross sections exhibit little sensitivity to the axial mass value and the calculations reproduce the data very well, and (ii) the dominant contribution to $\chi^2$ comes from the high-$\qrec$ region.


On the other hand, as shown in Table~\ref{tab:fits}, FSI effects turn out to be essential to reproduce the experimental cross sections of Refs.~\cite{ref:MINERvA_anu,ref:MINERvA_nu}. It is illustrated also in Fig.~\ref{fig:neutrinos}, where the numerical results accounting for FSI and neglecting them are represented by the solid and dashed lines, respectively. The axial mass applied in the calculations is set to 1.16 GeV.

The employed uncertainty estimate of the nuclear model is based on the extensive comparison
to electron-scattering data~\cite{ref:Barreau,ref:Whitney,ref:OConnell,ref:Baran,ref:Baghdasarian,ref:Sealock,ref:Day} performed in Ref.~\cite{ref:FSI}. It is important to note that it conservatively assumes that the cross section's uncertainty is 100\% at the kinematics corresponding to the excitation energy $E_x\lesssim26$ MeV, where the excitations of the giant dipole resonance and bound states contribute. Comparisons to low-energy data~\cite{ref:Czyz,ref:Ricco,ref:Yamaguchi,ref:Ryan}---left for future studies---can be expected to allow this assumption to be relaxed, reducing theoretical uncertainties and enabling much clearer discrimination of the calculations with and without FSI than that presented in Fig.~\ref{fig:neutrinos}.

As a final remark, I would like to acknowledge that results similar to these presented here have been reported from the relativistic Green's function model~\cite{ref:MINERvA_RGF}. This consistency is likely to be ascribed to the fact that the two approaches make use of the same optical potential~\cite{ref:Cooper_EDAI}. Unfortunately, the authors of Ref.~\cite{ref:MINERvA_RGF} have not pointed out an essential ingredient bringing their calculations into agreement with the MINERvA data.

In summary, a satisfactory description of the CC QE cross sections extracted from MINERvA can be achieved when FSI effects on the $\mu^\pm$ kinematics are accounted for in the nuclear model. It is important to keep in mind that the selection of the CC QE signal may exhibit a certain dependence on experimental details and is subject to systematic uncertainties stemming from backgrounds subtraction~\cite{ref:GiBUU_MINERvA}. Therefore, it is interesting to note that the effective values of the axial mass determined in this article are in rather good agreement with the result reported by the MiniBooNE Collaboration in Ref.~\cite{ref:MiniB_kappa}. This finding can be interpreted as pointing toward a non-negligible role of two-body reaction mechanisms at the kinematics of the MINERvA experiment.

I would like to thank Omar Benhar and Camillo Mariani for their comments on the manuscript.
This work has been supported by the National Science Foundation under Grant No. PHY-1352106.

\end{document}